\documentclass[cameraready]{vgtc}
\ifpdf%                                % if we use pdflatex
  \pdfoutput=1\relax                   % create PDFs from pdfLaTeX
  \usepackage{graphicx}                % allow us to embed graphics files
  \DeclareGraphicsExtensions{.pdf,.png,.jpg,.jpeg} % for pdflatex we expect .pdf, .png, or .jpg files
\else%                                 % else we use pure latex
  \usepackage{graphicx}                % allow us to embed graphics files
  \DeclareGraphicsExtensions{.eps}     % for pure latex we expect eps files
\fi%

\graphicspath{{figures/}{pictures/}{images/}{./}} % where to search for the images

\usepackage{microtype}                 % use micro-typography (slightly more compact, better to read)
\PassOptionsToPackage{warn}{textcomp}  % to address font issues with \textrightarrow
\usepackage{textcomp}                  % use better special symbols
\usepackage{mathptmx}                  % use matching math font
\usepackage{times}                     % we use Times as the main font
\usepackage{cite}                      % needed to automatically sort the references
\usepackage{tabu}                      % only used for the table example
\usepackage{booktabs}                  % only used for the table example
\usepackage{subfig}
\usepackage{svg}
\renewcommand\footnotemark{}

\usepackage{multirow}

\usepackage{float}

\usepackage{tabularx}

\usepackage{colortbl}
\usepackage{xcolor}
\usepackage{diagbox}  % Include this in your preamble
\usepackage[most]{tcolorbox}
\usepackage{hyperref}
\usepackage{cleveref}

\usepackage{makecell}  % Add this in your preamble

\vgtccategory{Research}

\vgtcinsertpkg

\title{Say It, See It: A Systematic Evaluation on Speech-Based 3D Content Generation Methods in Augmented Reality}

\author{
    Yanming Xiu$^1$ \\
    \scriptsize Department of Electrical and\\
    \scriptsize Computer Engineering \\
    \scriptsize Duke University \\
    \thanks{*Equal Contribution}
    \thanks{$^1$yanming.xiu@duke.edu}
    \and
    Joshua Chilukuri$^2$* \\
    \scriptsize North Carolina School of\\
    \scriptsize Science and Mathematics \\
    \scriptsize Durham, NC \\
    \thanks{$^2$joshua.chilukuri@gmail.com}
    \and
    Shunav Sen$^3$* \\
    \scriptsize North Carolina School of\\
    \scriptsize Science and Mathematics \\
    \scriptsize Durham, NC \\
    \thanks{$^3$shunavjainsen@gmail.com}
    \and
    Maria Gorlatova$^4$ \\
    \scriptsize Department of Electrical and\\
    \scriptsize Computer Engineering \\
    \scriptsize Duke University \\
    \thanks{$^4$maria.gorlatova@duke.edu}
}

\abstract{
As augmented reality (AR) applications increasingly require 3D content, generative pipelines driven by natural input such as speech offer an alternative to manual asset creation. In this work, we design a modular, edge-assisted architecture that supports both direct text-to-3D and text-image-to-3D pathways, enabling interchangeable integration of state-of-the-art components and systematic comparison of their performance in AR settings. Using this architecture, we implement and evaluate four representative pipelines through an IRB-approved user study with 11 participants, assessing six perceptual and usability metrics across three object prompts. Overall, text-image-to-3D pipelines deliver higher generation quality: the best-performing pipeline, which used FLUX for image generation and Trellis for 3D generation, achieved an average satisfaction score of 4.55 out of 5 and an intent alignment score of 4.82 out of 5. In contrast, direct text-to-3D pipelines excel in speed, with the fastest, Shap-E, completing generation in about 20 seconds. Our results suggest that perceptual quality has a greater impact on user satisfaction than latency, with users tolerating longer generation times when output quality aligns with expectations. We complement subjective ratings with system-level metrics and visual analysis, providing practical insights into the trade-offs of current 3D generation methods for real-world AR deployment.
} % end of abstract

\CCScatlist{
    \CCScatTwelve{Mixed / Augmented Reality}{Content Generation}{Evaluation Methods}{Quality Assessment};
}
\begin{document}


\maketitle

\section{Introduction}

Augmented Reality (AR) is becoming increasingly integrated into everyday digital experiences, from gaming and education to industrial and design applications~\cite{edu02, survey01, survey02}. As AR devices gain wider adoption, there is a growing demand for intuitive content creation tools that empower users to express ideas naturally and help developers streamline asset production through faster, more automated workflows. In practice, building an AR application or conducting AR-related research often requires significant time and effort to design, search for, or manually curate suitable 3D assets~\cite{bottleneck01}. This process can be a major bottleneck, especially when rapid prototyping or deployment is required. As a result, automating 3D content generation, particularly from simple inputs like speech or text, holds strong potential to accelerate development cycles and lower the barrier to entry for both developers and users.

Recent advances in generative AI, including speech-to-text, text-to-image, and image/text-to-3D models, provide a promising foundation for automated 3D content generation pipelines~\cite{dreammesh, minevra, matrix01, matrix02, matrix03, llmer}. These models open up new possibilities for enabling natural, hands-free content creation in AR. However, many existing methods still face practical limitations when deployed in real-time AR scenarios. For example, some pipelines require extensive computation time and resources, making them unsuitable for in-situ generation on resource-constrained devices. Others may produce low-fidelity or semantically misaligned outputs that fail to meet user expectations or blend naturally with the surrounding scene. Despite these challenges, there has been little effort to systematically evaluate the full generation pipeline in the AR context and from a system-level perspective. Existing studies tend to focus on either subjective user experience or isolated technical metrics, leaving a gap in understanding how these methods perform holistically in AR settings.

To address this gap, we present a comparative study of four speech-to-3D generation pipelines for AR applications. Our system incorporates a large language model (LLM) to refine transcribed user speech into optimized prompts, improving the robustness of downstream 3D content generation modules. We conduct a subjective user study to assess texture quality, content realism, mesh integrity, whether the content reflects the user's intent, and whether the users are annoyed  by the latency these pipelines require. In parallel, we evaluate each pipeline using objective system-level metrics, including generation latency and 3D asset file size. This dual-perspective evaluation highlights the strengths and limitations of existing approaches and provides insights into the design of more effective generative AR systems. The key contributions of this work include:

\begin{itemize}

    \vspace{-0.2cm} 

    % \item We design and evaluate four audio-to-3D generation pipelines, each integrating state-of-the-art components such as Whisper~\cite{whisper}, Mistral~\cite{mistral}, Shap-E~\cite{shap-e}, FLUX~\cite{flux},  TripoSR~\cite{TripoSR}, and Trellis~\cite{trellis}. These pipelines are deployed within a two-stage user-edge architecture and represent distinct designs for speech-driven 3D content generation. This modular structure enables a systematic comparison across alternative components.

    \item We introduce a modular, edge-assisted architecture for speech-driven 3D content generation in AR. The modular design allows interchangeable integration of alternative generation components, enabling a controlled comparison of different pipeline configurations. Within this architecture, we implement and evaluate four representative speech-to-3D pipelines that combine state-of-the-art modules, including Whisper~\cite{whisper}, Mistral~\cite{mistral}, Shap-E~\cite{shap-e}, FLUX~\cite{flux},  TripoSR~\cite{TripoSR}, and Trellis~\cite{trellis}, covering both direct text-to-3D and text-image-3D generation pathways. The pipelines are deployed on a Meta Quest 3 headset connected to an edge server.

    \vspace{-0.2cm} 

    \item We perform system-level evaluations to assess the capabilities of these pipelines, measuring generation latency and 3D asset file size, offering insights into the practical feasibility of each pipeline for AR applications with different latency and resource demands.

    \vspace{-0.2cm} 

    \item We conduct an IRB-approved user study to assess the subjective quality of generated content in AR scenes. Participants rated six aspects of the generated content across pipelines: texture quality, visual realism, mesh completeness, latency, user intent alignment and overall satisfactions, revealing key differences in user satisfaction. The results suggest that the strongest generation pipeline is capable of producing content that aligns well with user intent and meets user expectations. Although the generation latency may be long for real-time applications, user feedback indicates that it is generally acceptable when the output quality is high.

\end{itemize}
\section{Related Work}

Recent advancements in generative AI have enabled a range of pipelines for automatic 3D content creation from natural language or visual inputs. These pipelines typically combine modules such as speech-to-text, text-to-image, image-to-3D, or direct text-to-3D generation. Previous work has demonstrated the potential of these techniques in enabling intuitive content creation within AR and MR environments. Weng et al.~~\cite{dreammesh} proposed Dream Mesh, a speech-to-3D content generation system for mixed reality that integrates Whisper and DreamFusion, enabling users to generate 3D models through natural language prompts and interact with them in the MR environment. While this work demonstrates the potential of generative AI for spatial content creation, it does not include a user study, leaving the user experience unvalidated, and suffers from extremely long generation time: about 30 minutes per model generation, making it impractical for in-situ AR deployment. Santarnecchi et al.~\cite{minevra} proposed MineVRA, a VR system that generates 3D content through a fixed pipeline of speech-text-image-3D, incorporating environmental context for prompt refinement. They conducted a user study comparing generative results with Sketchfab assets, showing higher contextual coherence for AI-generated content. However, the system does not support direct text-to-3D generation and relies on image as intermediate step, resulting in a relatively high generation latency. The evaluation focuses only on user perception, with no system-level metrics, and the VR-only setup limits applicability to real-world AR use cases. Behravan et al.~\cite{matrix01, matrix02, matrix03} proposed a context-aware AR system that uses Vision Language Models (VLMs) like LLaVA~\cite{llava} to analyze captured scenes and recommend relevant 3D objects or recognize the existing objects in the real scene. These objects are then generated and integrated into the AR environment. However, the system does not support speech input and requires users to fill out structured text prompts, and the design paradigm restricts user creativity by only allowing the placement of ``context-reasonable" or existing objects rather than enabling open-ended 3D creation.

\begin{figure}[t]
\includegraphics[width=1.0\linewidth]{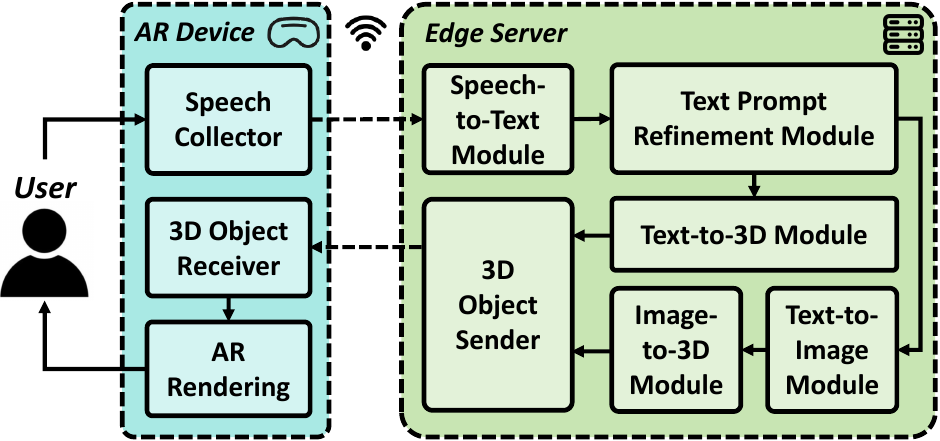}
\centering
\vspace{-0.3cm}
\caption{System architecture of our proposed speech-to-3D content generation pipeline for AR.}
\label{fig:diag}
\vspace{-0.6cm}
\end{figure}

In contrast to the previously discussed systems that generate new 3D content from user input, Chen et al.~~\cite{llmer} proposed LLMER, a framework that uses LLMs such as GPT-4 to generate structured JSON data for managing object creation, placement, and animation in XR environments. Rather than generating novel 3D models, LLMER focuses on controlling existing assets through modular execution, arguing that 3D generation is too time-consuming for real-time use. However, despite avoiding the cost of 3D model synthesis, the system relies on a cloud-based commercial LLM, which also introduces non-trivial latency that makes it unsuitable for real-time applications, and raises privacy concerns due to the use of non-local black-box services.

To address the limitations identified in prior work, our study focuses on systematically evaluating speech-driven 3D content generation pipelines in the context of AR. Unlike existing systems that usually rely on fixed architectures, subjective-only evaluation, or non-local commercial services, our approach emphasizes modular design, dual-perspective evaluation (subjective and objective), and full local deployment. This allows us to assess both the user experience and system performance of multiple generative pipeline designs while preserving user data privacy.

% However, most existing systems are designed as fixed pipelines and are rarely evaluated holistically. Key aspects such as latency, modularity, and user experience are often underexplored, making it difficult to assess their suitability for real-time, interactive AR use. In the following, we review representative systems that employ generative pipelines for immersive 3D content creation and analyze their strengths and limitations.

\begin{figure}[t]
\includegraphics[width=1.0\linewidth]{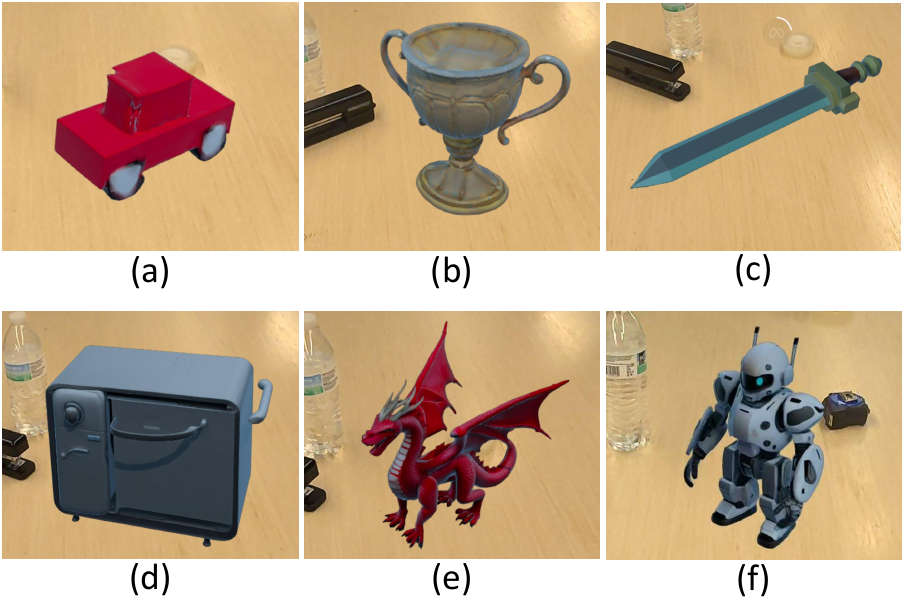}
\centering
\vspace{-0.6cm}
% \caption{Some examples of the 3D content generated by our audio-to-3D pipelines.
% (a): Prompt: ``Red toy car", created by Shap-E;
% (b): Prompt: ``Golden prize cup", created by FLUX +TripoSR;
% (c): Prompt: ``Medival style sword", created by Trellis (text mode);
% (d): Prompt: ``small microwave oven", created by Trellis (text mode);
% (e): Prompt: ``red fire dragon", created by FLUX + Trellis (image mode);
% (f): Prompt: ``Futuristic robot", created by FLUX + Trellis (image mode).
% }
\caption{Some examples of the 3D content generated by our speech-to-3D pipelines.
(a): ``Red toy car" by Shap-E;
(b): ``Golden prize cup" by FLUX + TripoSR;
(c): ``Medieval style sword" by Trellis (text mode);
(d): ``small microwave oven" by Trellis (text mode);
(e): ``red fire dragon" by FLUX + Trellis (image mode);
(f): ``Futuristic robot" by FLUX + Trellis (image mode).
}
\label{fig:output}
\vspace{-0.6cm}
\end{figure}

\section{Speech-to-3D Generation Pipelines}

While prior work has explored various pipelines for automatic 3D content generation~\cite{dreammesh, minevra, matrix01, matrix02}, most systems are built around fixed architectures or focus solely on either user experience or technical feasibility. Inspired by these efforts, we design a modular, edge-assisted pipeline that enables speech input-based 3D content creation tailored for AR applications. Our system supports multiple generation pathways, allowing for controlled comparisons across components, and emphasizes usability through local processing and efficient communication between the AR device and the server. As illustrated in Figure~\ref{fig:diag}, the architecture is divided into two main parts: the AR device and the edge server. This division allows the system to offload computation-intensive tasks while maintaining a responsive user experience.

\noindent \textbf{AR Device}: The AR device handles user interaction and content visualization. When a user speaks a prompt, the Speech Collector records the voice input and streams it to the edge server. Once a 3D model is generated, the 3D Object Receiver obtains it and passes it to the AR Rendering module, which renders the content into the user's view of the physical environment. This lightweight client-side setup enables deployment on resource-constrained AR headsets.

\noindent \textbf{Edge Server}: The edge server is responsible for the generative backbone of the system. The incoming speech audio from the user is first transcribed by the speech-to-text module, then passed through the text prompt refinement module, which hosts a light-weight LLM and reformulates the transcription into a clean, generation-friendly prompt. From here, the system supports two types of generation pipelines:

\begin{itemize}

    \vspace{-0.3cm}     
    
    \item A direct path, where the refined prompt is sent to the text-to-3D module without generating an image as an intermediate step.

    \vspace{-0.3cm} 
    
    \item A visual-intermediate path, where the prompt is passed to the text-to-image Module and then to the image-to-3D module.

    \vspace{-0.3cm} 
\end{itemize}

To realize this architecture, we use Whisper-Turbo~\cite{whisper} as the speech-to-text module and Mistral-7B~\cite{mistral} as the text prompt refinement module. The prompt for text prompt refinement is as follows:

\begin{tcolorbox}[colback=green!10!white, 
                  colframe=black, 
                  boxsep=4pt,  % inner padding
                  left=4pt,    % left padding
                  right=4pt,   % right padding
                  top=4pt,     % top padding
                  bottom=4pt   % bottom padding
                 ]
                 
``You need to clean and refine the given text. The text will contain a few words describing a certain object, and will be used for generating the corresponding image or 3D mesh that represents the object. Please remove all the unnecessary oral expressions and only leave the object and the descriptive words in the original text."

\end{tcolorbox}

\noindent With this design, noise in the recognized prompt is removed, leaving only the object description for generation. For example, if the user says, ``\textbf{\textit{Um, I think, please generate a red apple for me}}," the refined text prompt provided to the 3D content generation components will be simply ``\textbf{\textit{red apple}}," avoiding the influence of irrelevant words.

Based on the configurations above, we implement and compare four distinct generation pipelines, two of which generate 3D content directly from text prompts, while the other two follow a two-stage approach: first generating a 2D image from the prompt and then generating a 3D object from the image:

\noindent \textbf{1. Shap-E}: This pipeline employs Shap-E~~\cite{shap-e}, a well-known direct text-to-3D generation model developed by OpenAI. It has been widely adopted in prior implementations~\cite{dreammesh, matrix01, matrix02, octo+} of automatic 3D content generation for immersive applications.

\noindent \textbf{2. Trellis (Text Mode)}: This pipeline uses the text-conditioned version of Trellis~~\cite{trellis}, a recent high-capacity 3D generation model capable of synthesizing meshes from either textual or visual prompts. In this configuration, we provide only the refined text prompt for direct 3D synthesis.

\noindent \textbf{3. FLUX + TripoSR}: This pipeline has a two-stage structure, which uses FLUX.1-schnell~\cite{flux} as the text-to-image module, generating a 2D image from the text prompt. The resulting image is passed to TripoSR~\cite{TripoSR}, a recent model for 3D generation from a single image.

\noindent \textbf{4. FLUX + Trellis (Image Mode)}: Similar to the previous pipeline, this setup also uses FLUX.1-schnell for image generation. However, the image is then processed by the image-conditioned version of Trellis, which reconstructs a 3D asset based on visual input.

\begin{figure}[t]
\includegraphics[width=1.0\linewidth]{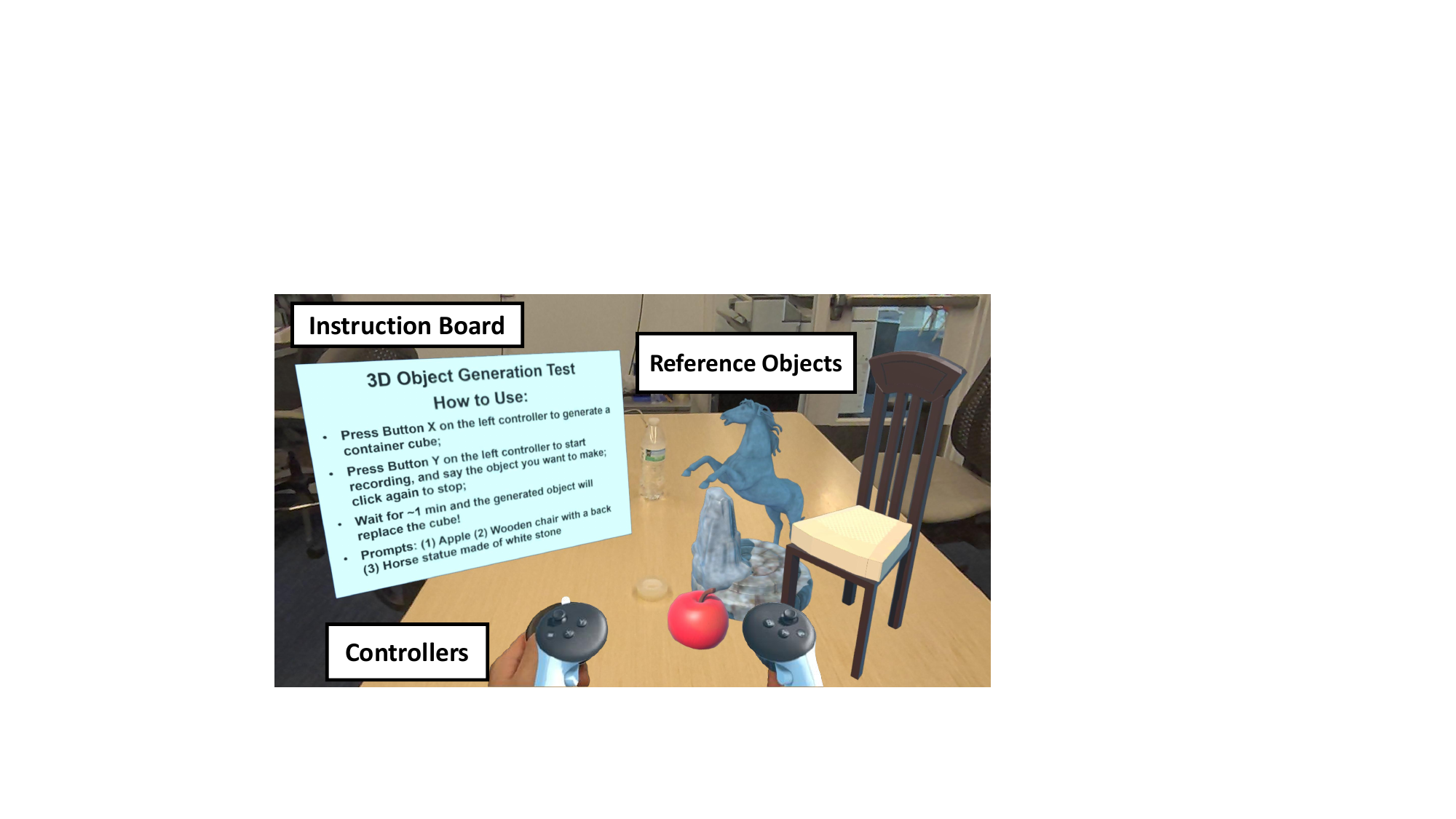}
\centering
\vspace{-0.4cm}
\caption{The scene setup of the AR application we developed for the user study. An instruction board and three reference objects (apple, chair, horse statue) are placed in the scene.}
\label{fig:study setup}
\vspace{-0.6cm}
\end{figure}

For the pipelines, the recorded speech input is transmitted as an .mp3 file, and intermediate images generated in the text-image-to-3D pathways are stored in .png format. The final 3D content is returned to the AR device in the .glb format, ensuring compatibility with standard AR rendering engines. All modules are containerized and independently callable, enabling modular experimentation and flexible pipeline orchestration. This setup allows us to conduct controlled benchmarking and head-to-head comparisons of various generation strategies under consistent runtime conditions, and avoids using any cloud-based services. Some examples of the pipelines' output are shown in Figure \ref{fig:output}. A video demonstrating our 3D generation pipelines can be accessed via this link.\footnote{\label{demo video}\url{https://youtube.com/watch/bCshDI_035c?feature=share}}

% Here we put the study setup

\section{System Evaluation}

\subsection{User Study Setup}

To evaluate the performance of our proposed 3D content generation pipelines, we conducted a user study across four generation pipelines. We developed an AR application for the user study, as shown in Figure~\ref{fig:study setup}. In the AR scene, an instruction board is provided to users to give them the necessary information to correctly use the app. Several 3D objects for comparison purposes are also placed in the scene. The user can trigger speech recording to start/stop using the controller. All of the 3D objects, including the instruction board, the reference objects and the generated objects are grabbable using the controller, giving users the freedom to manipulate the content's spatial properties. When recording starts or stops, the system plays audio cues of ``What should I make?" or ``Creating object" to help users better understand the current system state.

We then recruited 11 participants, aged from 18 to 30, for the user study, each of whom had some experience in AR technology, and four of whom were native English speakers. Each participant tested all four generation pipelines in a random order, using three fixed prompts representing different use cases: an ``\textbf{\textit{apple}}" (common object without description), a ``\textbf{\textit{wooden chair with a back}}" (common object with some description), and a ``\textbf{\textit{horse statue made of white stone}}" (uncommon object with rich description). The total duration of the study is about 30 minutes. The recognized speech input was monitored to ensure it matched the predefined standard prompts exactly, thereby eliminating the influence of noisy or inconsistent inputs, since evaluating the 3D generation components is our primary objective. For each generated object, users were asked to evaluate the output through a 6-question Likert-style questionnaire~\cite{likert} covering: texture quality, visual realism, mesh completeness, latency annoyance, alignment with user intent, and overall satisfaction. The questions are listed below:

\begin{tcolorbox}[colback=green!10!white, 
                  colframe=black, 
                  boxsep=3pt,  % inner padding
                  left=4pt,    % left padding
                  right=4pt,   % right padding
                  top=4pt,     % top padding
                  bottom=4pt   % bottom padding
                 ]
                 
\noindent \textbf{Q1}: How do you feel about the texture quality of the generated [object]? 1 for very poor and 5 for very good. Note: you can consider the reference [object] as 3.

\noindent  \textbf{Q2}: How do you feel about the object realism of the generated [object]? 1 for very poor and 5 for very good. Note: you can consider the reference [object] as 3.

\noindent \textbf{Q3}: Does the generated [object] have a complete and intact mesh (no visible holes, gaps, or noise particles)? 1 for many artifacts and 5 for no artifacts.

\noindent \textbf{Q4}: Does the latency for generating the [object] make you feel bothered or annoyed? 1 for very annoyed and 5 for not at all.

\noindent \textbf{Q5}: Does the generated [object] align with your intent / imagination? 1 for very poorly aligned and 5 for very well aligned.

\noindent \textbf{Q6}: Overall, how do you like the [object] generated by this pipeline? 1 for dislike very much and 5 for like very much.

\end{tcolorbox}

% Here we put all the generated content
\begin{figure*}[t]
\includegraphics[width=1.0\linewidth]{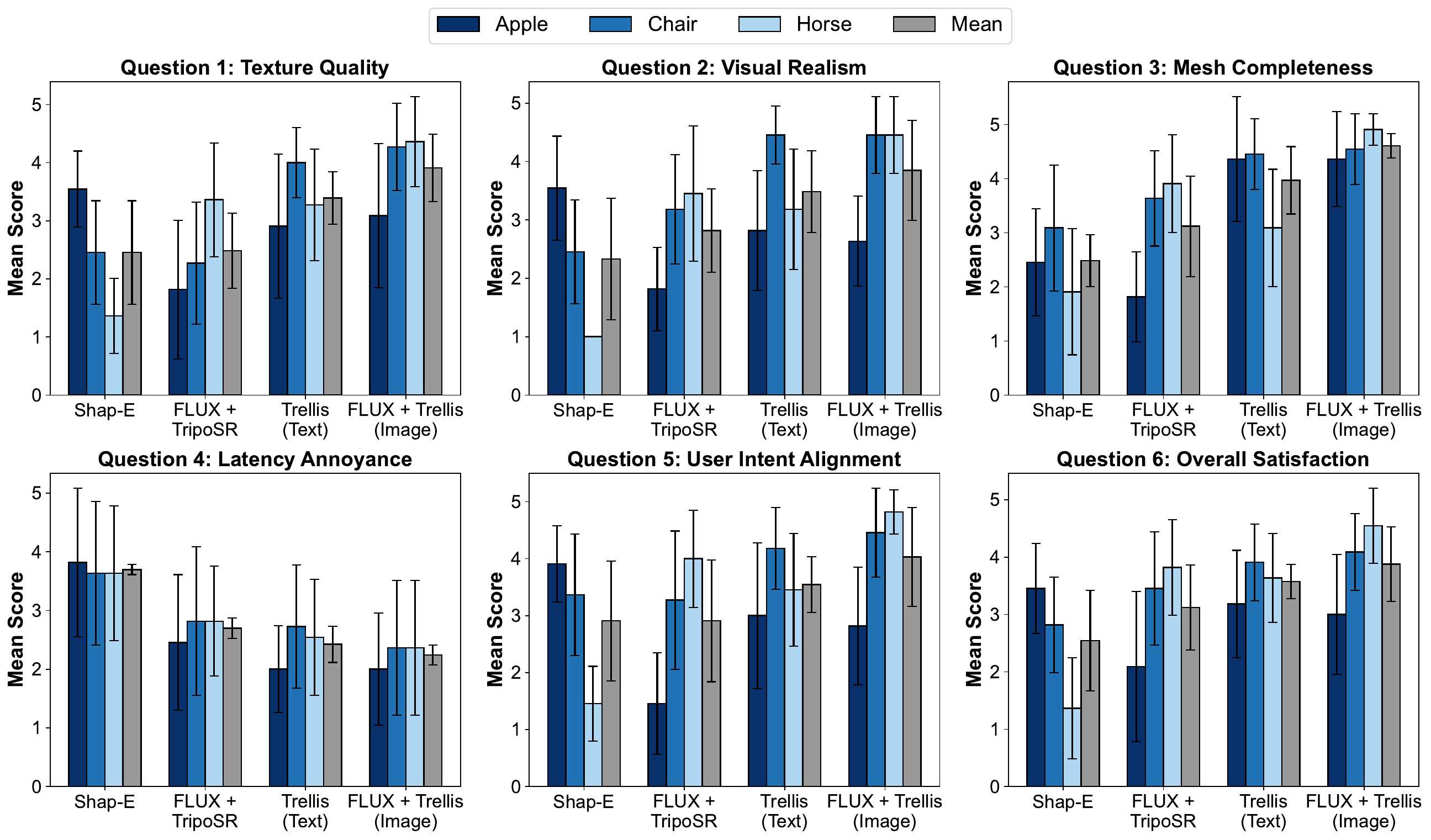}
\centering
\vspace{-0.5cm}
\caption{User study results for the 5-point Likert scale scores of six evaluation questions across four generation pipelines and three object types, including the average score.}
\label{fig:res1}
\vspace{-0.1cm}
\end{figure*}

\begin{table*}[ht]
\centering
\footnotesize
\caption{System-level comparison of 3D generation pipelines across different prompts. }
\vspace{-0.2cm}
\begin{tabular}{l|ccc|ccc|ccc}
\toprule
\textbf{\makecell{3D Content Generation Pipeline}} & \multicolumn{3}{c|}{\textbf{Generation Latency (s)}} & \multicolumn{3}{c|}{\textbf{File Size (MB)}} & \multicolumn{3}{c}{\textbf{Overall Satisfaction Level (Q6)}}\\
 & Apple & Chair & Horse & Apple & Chair & Horse & Apple & Chair & Horse\\
\midrule
\makecell{Shap-E} & \textbf{21.19} & \textbf{17.76} & \textbf{18.31} & \textbf{0.19} & \textbf{0.19} & \textbf{0.19} & \textbf{3.45} & 2.82 & 1.36\\
\midrule
\makecell{FLUX + TripoSR} & 37.09 & 32.73 & 32.72 & 3.79 & 1.37 & 1.39 & 2.09 & 3.45 & 3.82\\
\midrule
\makecell{Trellis (Text)} & 65.06 & 33.44 & 39.08 & 1.04 & 1.23 & 1.54 & 3.18 & 3.91 & 3.64\\
\midrule
\makecell{FLUX + Trellis (Image)} & 74.23 & 50.55 & 50.03 & 1.24 & 1.37 & 1.42 & 3.00 & \textbf{4.09} & \textbf{4.55} \\
\bottomrule
\end{tabular}

\vspace{-0.3cm}
\label{tab:latency_file_size}
\end{table*}

 Among the questions, Q1, Q2, and Q3 primarily reflect the perceptual and semantic fidelity of the generated content, while Q4 captures system-level concerns. Lastly, Q5 and Q6 evaluate the users' overall subjective preference. We notice that Q1 and Q2 assess the perceptual quality of the generated content, including its texture fidelity and overall realism, which are both inherently subjective. To help users make more grounded evaluations, we placed three reference objects (apple, wooden chair, and horse statue) within the AR scene. These reference objects exhibit moderate levels of texture detail and realism, serving as visual baselines during the study.

In addition to subjective ratings, we also collected objective system-level metrics, including (1) generation latency, which records the time between when the user stops speaking and when the generated virtual content is rendered in the headset; and (2) the file size of the generated 3D content for each object-pipeline combination. This dual evaluation setup enables a comprehensive comparison of both perceived and technical performance of the generation pipelines under realistic AR conditions.

We deploy the AR interface on a Meta Quest 3, while all computation-intensive modules are hosted on an edge server equipped with three NVIDIA RTX 3090 GPUs. The AR application is developed using Unity 6000.1.7f1, and data exchange between the AR device and the server is facilitated by a single-hop 5GHz WiFi connection, ensuring low-latency communication and responsive interaction.

\begin{figure*}[t]
\includegraphics[width=1.0\linewidth]{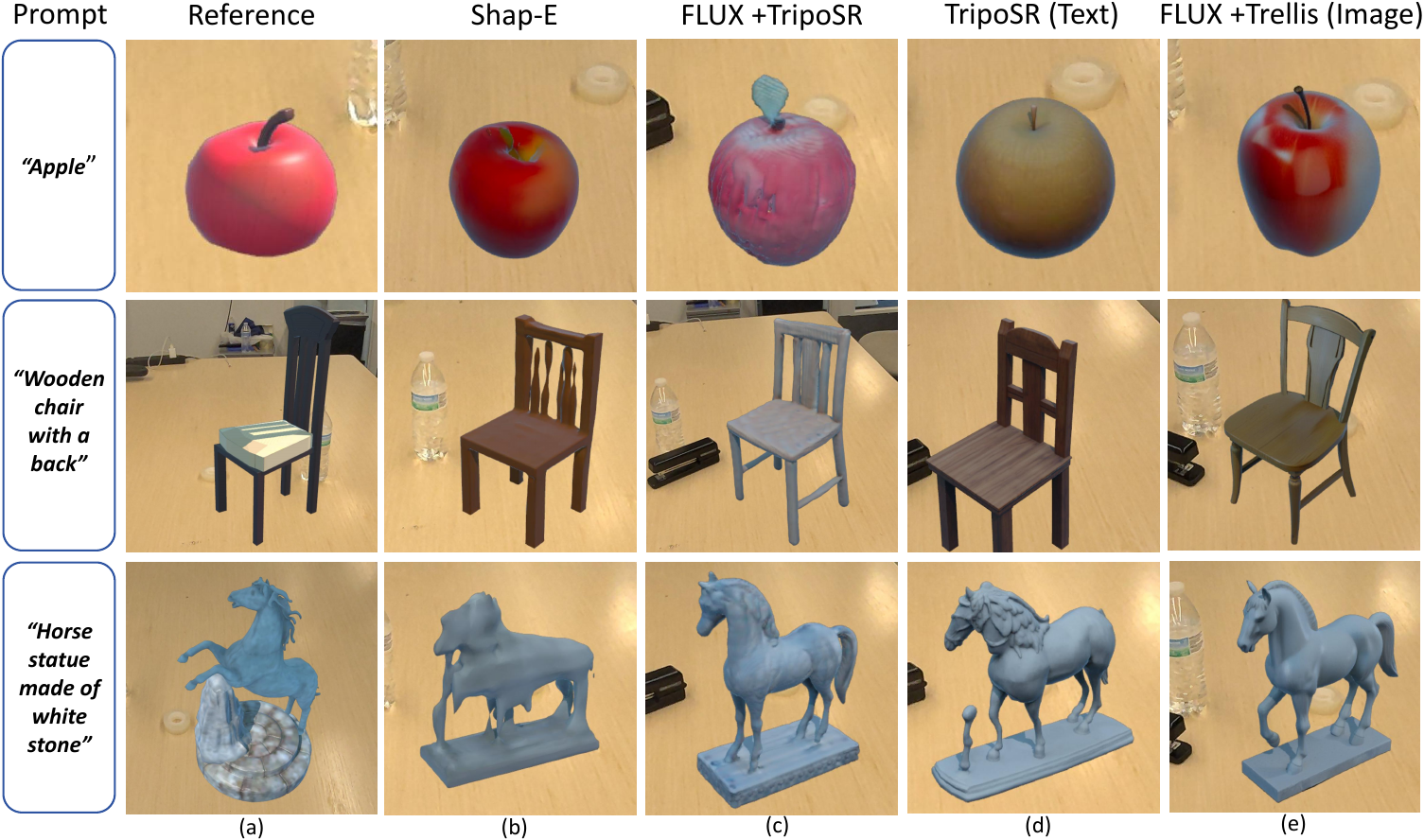}
\centering
\vspace{-0.6cm}
\caption{Visual comparison of generated 3D content for three object categories: apple (top row), wooden chair with a back (middle row), and horse statue made of white stone (bottom row). (a): Reference objects with moderate quality. Specifically, the apple has a well-formed mesh but an unrealistic texture; the chair features a conventional yet slightly unnatural structure; and the horse statue has a realistic texture but awkward overall proportions of the head, body, and legs. (b)-(e) objects generated by Shap-E, FLUX + TripoSR, Trellis and FLUX + Trellis, respectively. Several real-world elements (e.g., water bottle, tape) are added to the AR scene to provide context for scale, realism, and visual integration. Generally, FLUX + Trellis has the best performance, while it may still perform poorly in certain tasks, such as the texture quality of apple generation.}
\label{fig:vis}
\vspace{-0.5cm}
\end{figure*}

\subsection{User Study Results Analysis}

The results of our user study are shown in Figure~\ref{fig:res1}. Besides, we also monitor the pipelines' latency in generating these content as well as the file size of the content, as shown in Table~\ref{tab:latency_file_size}. To understand the performance differences among the four generation pipelines, we analyze the results based on the three main aspects: perceptual quality, system-level performance, and user satisfaction and alignment with intent, as mentioned in the questionnaire design. 

\noindent \textbf{Perceptual Quality}: Across Q1-Q3, we observe a general consistent trend in which the FLUX + Trellis pipeline outperforms the others with average scores of 3.91, 3.85, and 4.03 for Q1, Q2, and Q3, respectively, followed by Trellis (Text), FLUX + TripoSR, and finally Shap-E. This ranking is especially pronounced in the generation of creative or uncommon objects, such as the horse statue, where the gap between pipelines becomes more significant. The superior performance of FLUX + Trellis likely stems from the complementary strengths of both components: the image generation module mitigates limitations in text-only descriptions, while Trellis provides high-fidelity 3D reconstructions.

For more common objects such as the apple or chair, the gap between Trellis and FLUX + Trellis narrows. This is likely due to the prevalence of these object categories in the training data of modern text-to-3D models, which allows Trellis to perform competitively even without intermediate image guidance. Notably, Shap-E, despite being the earliest proposed and least capable overall, performs surprisingly well on the apple task, outperforming all three other pipelines. This can be attributed to its extensive training on simple, commonly occurring object classes, in contrast to its poor generalization on complex or less typical prompts like ``horse statue". These results indicate that as for the current pipelines, intermediate image guidance substantially improves the realism and completeness of generated assets, especially for uncommon or complex prompts.

\noindent \textbf{System-Level Performance}: In terms of latency and file size, Shap-E consistently outperforms all other pipelines, achieving generation times of approximately 20 seconds and producing compact assets under 0.2 MB across all object types. This efficiency is further supported by user responses to Q4, which reflect lower perceived annoyance with Shap-E’s latency. Overall, we observe an inverse relationship between generation latency and perceptual quality. Notably, the inclusion of an image generation stage in the FLUX + Trellis pipeline increases latency by roughly 10 seconds compared to the text-only Trellis pipeline. This highlights a key trade-off: for latency-sensitive applications, direct text-to-3D pipelines may be preferable, whereas quality-focused applications can justify the additional delay introduced by image-based pathways.

\noindent \textbf{User's Intent Alignment and Satisfaction}: In terms of alignment with user intent (Q5) and overall satisfaction (Q6), FLUX + Trellis (Image) consistently receives the highest scores (4.03 for intent alignment and 3.88 for overall rating), particularly for the horse statue generation task (4.82 for intent alignment and 4.55 for overall rating). This underscores its advantage in bridging the semantic gap between user descriptions and generated 3D content. Trellis (Text) also performs well, especially for familiar objects like chair, but begins to falter on uncommon prompts without the aid of visual grounding. Conversely, Shap-E, while efficient in latency and file size, shows the weakest performance on Q5 and Q6. This reflects user frustration with poor semantic fidelity and low content quality, even when generation is fast. Notably, user feedback suggests that overall satisfaction level is more sensitive to generation quality than speed. According to users' post-study feedback, the difference in generation latency between 20 seconds and 60 seconds is tolerable, as long as the output aligns with user expectations and looks realistic. Taken together, these patterns support our main conclusion that perceptual quality has a greater influence on user satisfaction than generation speed.

\noindent \textbf{Statistical Validation}: Finally, to statistically validate the observed trends, we performed Friedman tests on all user ratings across the four pipelines for each question and object. The results revealed significant differences in scores for all six questions across all object types (p $<$ 0.05), confirming that different structures of the generation pipelines significantly affect perceptual quality, system-level experience, intent alignment, and overall satisfaction.

\subsection{Visual-Based Qualitative Evaluation}

While the preceding analysis presents quantitative and user study results, we would like to offer a more intuitive understanding of the generation quality through visual inspection of representative outputs. Figure~\ref{fig:vis} showcases some of the generated 3D objects during user studies, for three different prompts produced by the four evaluated pipelines.

Among the four pipelines, Shap-E exhibits the lowest quality, often generating meshes with noticeable flaws in geometry, realism, and texture fidelity. Although it performs reasonably well for the apple prompt, likely due to the simplicity and prevalence of such objects in its training data, it fails to maintain quality for more complex or less common objects. This aligned with our findings in the user study. The generated chair lacks realistic surface texture, and the horse statue suffers from serious mesh deformation.

FLUX + TripoSR pipeline improves upon Shap-E in terms of mesh structure and object coherence, producing more stable shapes overall. However, realism and texture quality remain significant issues, particularly for objects that require nuanced material representation. For example, while the generated chair form is more consistent than Shap-E’s, the surface lacks the detail of wooden material and fails to blend naturally into the environment.

Trellis (Text) and FLUX + Trellis (Image) both demonstrate strong capabilities in generating realistic and visually coherent 3D content. This trend has been confirmed in our user study: the overall satisfaction scores of these two pipelines are higher than those of the other two. Nevertheless, some issues remain. The horse statue generated by Trellis (text) shows an ``extra leg" in the mesh geometry, affecting its structural plausibility. Similarly, the apple generated by FLUX + Trellis (Image) appears visually appealing from the front but lacks color on the back side, revealing challenges in complete surface texturing. 

Overall, these qualitative examples help to illustrate specific strengths and weaknesses of each pipeline. They complement and supplement our quantitative findings by providing intuitive visual evidence that supports the trends observed in the user study.
\section{Discussion}

Our study demonstrates that modern 3D generation pipelines, when integrated into AR systems, are capable of producing assets that rival or even surpass moderately detailed handcrafted models in perceptual quality while offering a nearly real-time feedback experience to users. This highlights the growing potential of generative AI to assist in rapid AR content creation, especially for developers and creators without access to professional asset design resources. Among the four evaluated pipelines, those incorporating an image intermediate step, particularly FLUX + Trellis, consistently outperformed direct text-to-3D methods on uncommon or complex prompts. This performance advantage suggests that inserting an image generation stage helps bridge the semantic and visual gap between textual descriptions and 3D geometry, providing more guidance for difficult or ambiguous prompts. The intermediate image acts as a visual scaffold, offering spatial structure and texture cues that may not be explicit in the text.

On the other hand, direct pipelines such as Trellis (text only) also showed some capability on simple or commonly encountered objects, such as ``apple" or ``wooden chair with a back." This result reflects both their architectural efficiency and the likelihood that these object categories are well represented in their training datasets. These findings suggest that with more diverse training data and better pretraining strategies, direct text-to-3D pipelines may also generate high-quality virtual content while largely reducing the content generation latency. 

Notably, while generation latency remains a key bottleneck for real-time or interactive AR applications, our user feedback indicates that perceptual quality has a more substantial influence on overall satisfaction. Most participants considered generation delays around 60 seconds tolerable when the resulting object matched their mental image and appeared realistic within the scene. In contrast, faster pipelines that produced lower-fidelity or semantically misaligned assets were often rated poorly despite their short wait times. This trade-off reveals that for many creative or content-authoring AR use cases, users are likely to tolerate modest delays if it leads to significantly better results, making quality-focused pipelines a practical choice despite current computational costs. Some participants further noted that their tolerance for such delays stemmed from prior experience in manually creating or searching for 3D assets, which are often far more time-consuming than the evaluated pipelines, and cannot intuitively preview the assets in the real world.

\section{Limitations and Future Work}

While our system demonstrates the feasibility of evaluating modular speech-driven 3D content generation pipelines for AR, several limitations point to natural directions for future work.

First, our current evaluation is limited to predefined text prompts and object categories. While useful for controlled system comparison and evaluation, this setting may not fully reflect how users interact with generative systems in real AR applications. In future work, we plan to expand the study to include open-ended generation tasks, where users describe and create objects of their own choosing. This will allow for a more authentic assessment of semantic alignment and user satisfaction.

Second, although our user study covers six perceptual and usability criteria across four pipelines and three object types, the number of participants, types of tasks and formats of users' feedback remains limited. We will address this by conducting a larger-scale user study involving more participants, more diverse prompts, additional models, and more open-ended questions, thereby improving the statistical power and generalizability of the results. Currently, our pipelines are built entirely with open-source models, primarily to avoid the privacy and cost concerns associated with cloud-based services. However, in scenarios where privacy is less critical, certain components such as the text-to-image module could be substituted with more powerful commercial alternatives like OpenAI's GPT~\cite{GPT4o} or Google's Gemini~\cite{comanici2025gemini}. These models may help to enhance the overall 3D generation quality, offering a potential direction for future system upgrades. We are also designing open-ended questions such as ``What do you like about the generated 3D content?"/``What aspect of the generated 3D content makes you frustrated", to capture richer qualitative feedback. Thematic analysis of such responses will allow us to identify recurring user expectations and pain points that may not be fully reflected in numerical scores, providing deeper insight into how to improve generation quality and user experience.

Finally, our current system operates in a testing context and is not yet deployed within real AR applications. To move toward practical utility, we plan to integrate the generation pipelines into functional AR applications. In particular, we will explore educational tools~\cite{edu01} that allow students to generate and manipulate learning-related 3D content via speech, as well as our recent work on AR safety applications that simulate visual attack scenarios in AR by automatically generating obstructive or misleading virtual content~\cite{viddar, vimsense}. 

By combining modular pipeline design with scalable evaluation methods, we hope this work lays the foundation for more intelligent, responsive, and user-centered 3D content generation in AR. As generative models continue to evolve, we will work on their integration into real-world applications and open up new possibilities for more immersive, personalized, and responsive AR experiences.

\acknowledgments{
Two coauthors, Joshua Chilukuri and Shunav Sen, completed this work through the North Carolina School of Science and Mathematics (NCSSM) Mentorship Program, which connects NCSSM students with external research mentors; we appreciate the program’s coordination and support. We thank the participants of our user study for their effort in this research. This work was supported in part by NSF grants CSR-2312760, CNS-2112562, and IIS-2231975, NSF CAREER Award IIS-2046072, NSF NAIAD Award 2332744, a Cisco Research Award, a Meta Research Award, Defense Advanced Research Projects Agency Young Faculty Award HR0011-24-1-0001, and the Army Research Laboratory under Cooperative Agreement Number W911NF-23-2-0224. The views and conclusions contained in this document are those of the authors and should not be interpreted as representing the official policies, either expressed or implied, of the Defense Advanced Research Projects Agency, the Army Research Laboratory, or the U.S. Government. This paper has been approved for public release; distribution is unlimited. No official endorsement should be inferred. The U.S.~Government is authorized to reproduce and distribute reprints for Government purposes notwithstanding any copyright notation herein.
}

\bibliographystyle{abbrv-doi}
%\bibliographystyle{abbrv-doi-narrow}
%\bibliographystyle{abbrv-doi-hyperref}
%\bibliographystyle{abbrv-doi-hyperref-narrow}

% \bibliography{template}

\begin{thebibliography}{10}

\bibitem{edu01}
C.~Avila-Garzon, J.~Bacca-Acosta, J.~Duarte, J.~Betancourt, et~al.
\newblock Augmented reality in education: An overview of twenty-five years of research.
\newblock {\em Contemporary Educational Technology}, 13(3), 2021.

\bibitem{matrix03}
M.~Behravan and D.~Gracanin.
\newblock Generative multi-modal artificial intelligence for dynamic real-time context-aware content creation in augmented reality.
\newblock In {\em Proceedings of the 30th ACM Symposium on Virtual Reality Software and Technology (VRST)}, 2024.

\bibitem{matrix02}
M.~Behravan, M.~Haghani, and D.~Gračanin.
\newblock Transcending dimensions using generative {AI}: Real-time 3{D} model generation in augmented reality.
\newblock In {\em International Conference on Human-Computer Interaction}, 2025.

\bibitem{matrix01}
M.~Behravan, K.~Matković, and D.~Gračanin.
\newblock Generative {AI} for context-aware 3{D} object creation using vision-language models in augmented reality.
\newblock In {\em 2025 IEEE International Conference on Artificial Intelligence and eXtended and Virtual Reality (AIxVR)}, 2025.

\bibitem{llmer}
J.~Chen, X.~Wu, T.~Lan, and B.~Li.
\newblock {LLMER}: Crafting interactive extended reality worlds with {JSON} data generated by large language models.
\newblock {\em IEEE Transactions on Visualization and Computer Graphics}, 31(05), 2025.

\bibitem{comanici2025gemini}
G.~Comanici, E.~Bieber, M.~Schaekermann, I.~Pasupat, N.~Sachdeva, I.~Dhillon, M.~Blistein, O.~Ram, D.~Zhang, E.~Rosen, et~al.
\newblock Gemini 2.5: Pushing the frontier with advanced reasoning, multimodality, long context, and next generation agentic capabilities.
\newblock {\em arXiv preprint arXiv:2507.06261}, 2025.

\bibitem{survey02}
S.~Dargan, S.~Bansal, M.~Kumar, A.~Mittal, and K.~Kumar.
\newblock Augmented reality: A comprehensive review.
\newblock {\em Archives of Computational Methods in Engineering}, 30(2), 2023.

\bibitem{GPT4o}
A.~Hurst, A.~Lerer, A.~P. Goucher, A.~Perelman, A.~Ramesh, A.~Clark, A.~Ostrow, A.~Welihinda, A.~Hayes, A.~Radford, et~al.
\newblock {GPT-4o} system card.
\newblock {\em arXiv preprint arXiv:2410.21276}, 2024.

\bibitem{mistral}
A.~Q. Jiang, A.~Sablayrolles, A.~Mensch, C.~Bamford, D.~S. Chaplot, D.~de~las Casas, F.~Bressand, G.~Lengyel, G.~Lample, L.~Saulnier, L.~R. Lavaud, M.-A. Lachaux, P.~Stock, T.~L. Scao, T.~Lavril, T.~Wang, T.~Lacroix, and W.~E. Sayed.
\newblock Mistral 7b, 2023.

\bibitem{shap-e}
H.~Jun and A.~Nichol.
\newblock Shap-{E}: Generating conditional 3d implicit functions, 2023.

\bibitem{flux}
B.~F. Labs.
\newblock Flux.
\newblock \url{https://github.com/black-forest-labs/flux}, 2024.

\bibitem{likert}
R.~Likert.
\newblock A technique for the measurement of attitudes.
\newblock {\em Archives of {P}sychology}, 1932.

\bibitem{bottleneck01}
J.~Linares-Pellicer, J.~Izquierdo-Domenech, I.~Ferri-Molla, and C.~Aliaga-Torro.
\newblock Breaking the bottleneck: Generative {AI} as the solution for {XR} content creation in education.
\newblock In {\em Advanced Technologies and the University of the Future}, pp. 9--30. Springer, 2024.

\bibitem{llava}
H.~Liu, C.~Li, Q.~Wu, and Y.~J. Lee.
\newblock Visual instruction tuning.
\newblock In {\em Proceedings of NeurIPS}, vol.~36, 2023.

\bibitem{whisper}
A.~Radford, J.~W. Kim, T.~Xu, G.~Brockman, C.~McLeavey, and I.~Sutskever.
\newblock Robust speech recognition via large-scale weak supervision, 2022.

\bibitem{minevra}
E.~Santarnecchi, E.~Balloni, M.~Paolanti, E.~Frontoni, L.~Stacchio, P.~Zingaretti, and R.~Pierdicca.
\newblock {MineVRA}: Exploring the role of generative {AI}-driven content development in {XR} environments through a context-aware approach.
\newblock {\em IEEE Transactions on Visualization and Computer Graphics}, 31(05), 2025.

\bibitem{octo+}
A.~Sharma, L.~Yoffe, and T.~Höllerer.
\newblock {OCTO}+: A suite for automatic open-vocabulary object placement in mixed reality.
\newblock In {\em IEEE International Conference on Artificial Intelligence and eXtended and Virtual Reality (AIxVR)}, 2024.

\bibitem{survey01}
Y.~Siriwardhana, P.~Porambage, M.~Liyanage, and M.~Ylianttila.
\newblock A survey on mobile augmented reality with 5{G} mobile edge computing: Architectures, applications, and technical aspects.
\newblock {\em IEEE Communications Surveys and Tutorials}, 23(2), 2021.

\bibitem{TripoSR}
D.~Tochilkin, D.~Pankratz, Z.~Liu, Z.~Huang, , A.~Letts, Y.~Li, D.~Liang, C.~Laforte, V.~Jampani, and Y.~Cao.
\newblock Tripo{SR}: Fast {3D} object reconstruction from a single image.
\newblock {\em arXiv preprint arXiv:2403.02151}, 2024.

\bibitem{dreammesh}
S.~C. Weng, Y.~Chiou, and E.~Y. Do.
\newblock {Dream Mesh}: A speech-to-{3D} model generative pipeline in mixed reality.
\newblock In {\em 2024 IEEE International Conference on Artificial Intelligence and eXtended and Virtual Reality (AIxVR)}, 2024.

\bibitem{trellis}
J.~Xiang, Z.~Lv, S.~Xu, Y.~Deng, R.~Wang, B.~Zhang, D.~Chen, X.~Tong, and J.~Yang.
\newblock Structured {3D} latents for scalable and versatile {3D} generation.
\newblock {\em arXiv preprint arXiv:2412.01506}, 2024.

\bibitem{vimsense}
Y.~Xiu and M.~Gorlatova.
\newblock Detecting visual information manipulation attacks in augmented reality: a multimodal semantic reasoning approach.
\newblock {\em arXiv preprint arXiv:2507.20356}, 2025.

\bibitem{viddar}
Y.~Xiu, T.~Scargill, and M.~Gorlatova.
\newblock {ViDDAR}: Vision language model-based task-detrimental content detection for augmented reality.
\newblock {\em IEEE Transactions on Visualization and Computer Graphics}, 31(05), 2025.

\bibitem{edu02}
F.~Zulfiqar, R.~Raza, M.~O. Khan, M.~Arif, A.~Alvi, and T.~Alam.
\newblock Augmented reality and its applications in education: A systematic survey.
\newblock {\em IEEE Access}, 11, 2023.

\end{thebibliography}

\end{document}